# Investigating Technological Solutions for Addressing Water Scarcity in Agricultural Production

Ji Woo Han

**[Abstract]**

This comprehensive study investigates the intricate relationship between water scarcity and agricultural production, emphasizing its critical global significance. The research, through multidimensional analysis, investigates the various effects of water scarcity on crop productivity, especially the economic water scarcity (AEWS) which is the main factor of influence. The study stresses the possibility of vertical farming as a viable solution to the different kinds of water scarcity problems, hence, it emphasizes its function in the sustainable agricultural development. Although the study recognizes that some problems still remain, it also points out the necessity of more research to solve the issues of scalability and socio-economic implications. Moving forward, interdisciplinary collaboration and technological innovation are essential to achieving water-secure agriculture and societal resilience.

# 1. [Introduction]

In the complex network of world problems that faces humanity, few issues come with the menacing gravity of water shortage. As we are on the verge of environmental destruction and the widening of social inequality, the necessity to deal with this problem becomes more and more obvious. The dark picture of water shortage is stretched over the areas of agricultural productivity, economic resilience and the basis of global food security, thus painting a gloomy picture of a future where access to this basic resource is not a matter of certainty. Water scarcity is not just a local problem; it is a complicated, intertwined issue that echoes across boundaries and influences communities globally (Ling, 2022). In other words, it is a clear representation of the fact that the increase in the demand for water and the limited nature of freshwater resources are not in agreement. Among the factors that make the problem of water supply more complicated are population growth, urbanization, industrialization, and climate change, which raise the demand for water and put a lot of pressure on the water resources and the infrastructure that supports them. In the agricultural areas, which are the water-based heart of the crops and livestock, the water resources are decreasing, which in turn threatens the livelihoods and food production. Farmers are the ones who are facing the reality of the shrinking of the irrigation supplies and the unpredictable rainfall patterns, which makes them have to choose between sustaining their fields and preserving the water for the essential needs. The consequences of the decrease in agricultural production go beyond the food system and affect the whole world, thus, making food insecurity and the increase of prices even more severe, especially in areas where poverty and malnutrition are already existing problems. Besides, water scarcity is a factor that affects the economic stability, because industries that are dependent on water for manufacturing, energy production, and resource extraction are faced with increasing difficulties (Omolere, 2024). Water scarcity is the main cause of the disruptions in the supply chains, which in turn, reduces productivity and increases the operational costs, thus, it is a serious threat to the local and global economies. In places where water shortage is combined with the existing social inequalities, the marginalized communities are the ones who feel the most of the pressure, and they face the higher chances of poverty, displacement, and conflict. The problem of water shortage is so important that it cannot be underestimated. It is a complex issue that involves the water management practices, the investments in water infrastructure, the technological innovations, and the equitable governance frameworks. The most important aspect of the extension of the limited water supplies is the conservation of water, and the efforts such as the improvement of irrigation efficiency and the promotion of water-saving technologies are the main factors of the conservation. Moreover, the promotion of the integrated water resource management strategies that are based on the cooperation and coordination of the stakeholders can be the way to optimize the water allocation and to reduce the conflicts that are caused by the competing water demands.

My motivation to delve into the investigation of technological solutions for addressing water scarcity in agricultural production stemmed from a deeply resonant source: The portrayal of dystopian futures in narratives like "The Silent Sea." The Korean drama, set in a world that is dealing with the consequences of the depletion of resources and social inequality, is a warning of the danger of the water resources and the disaster of their mismanagement (Roth, 2022). The characters' fight against the severe water scarcity, their use of extreme measures to get this vital resource, the vivid emotions that this evokes, all go beyond the limits of the fiction. The picture of a planet under the attack of environmental pollution and social conflicts makes me want to find the actual solutions to the problem of water shortage, which is the most important issue of our time. Besides its entertainment aspect, "The Silent Sea" highlights the need to deal with water scarcity, especially in the case of agricultural production. It emphasizes the relationship between water supply, food security, and social stability, thus making people think about the consequences of doing nothing when it comes to this urgent issue. Through this narrative and its thematic exploration of water scarcity, I was motivated to study technological innovation more closely. I was aware of the important role that technology can have in water efficiency, the irrigation practices and the mitigation of water scarcity impacts on agricultural yields.

      This paper tries to untangle the complex web of the world water scarcity story, looking at its root causes, its wide-ranging effects and the ways to get out of the situation and to build a sustainable future. I will use an interdisciplinary approach to overcome the limits of the traditional language, combining the knowledge from the fields of hydrology, agronomy, economics, and political science. Through the integration of these different parts of the knowledge, I will be able to give a complete view of the water scarcity that is not limited to any discipline and will be the basis for the collaborative solutions. The main aim of this research is to realize that water scarcity is not only a problem of the environment or humanity but also a vital issue that threatens the survival and the prosperity of human civilization. As agriculture is the biggest consumer of the world's water resources, the development of new technology that will improve water efficiency and agricultural productivity is now more urgent than ever. The technological breakthroughs in the fields of precision irrigation systems and drought-resistant crop varieties are the opportunities to change the way we manage water in agriculture and other places. The main element of my question is a thorough examination of the current literature which provides the necessary information on the complicated relationship of the factors that cause water scarcity and the success of the interventions made to solve the problem. Studies such as "Water Scarcity in Agriculture: The "An Overview of Causes, Impacts, and Approaches for Reducing the Risks" gives a complete view of the problems created by water scarcity and stresses the necessity of stakeholder involvement and collective action in the formulation of the water management strategies. Likewise, "Vertical Farming - Smart Urban Agriculture for Enhancing Resilience and Sustainability in Food Security" discusses the possibility of vertical farming as a food security solution, pointing out the necessity of increasing the output and reducing the resource inputs. Building upon these foundational works, my research endeavors to address the central question: Which technological solutions are considered as the best for the improvement of agricultural productivity in the face of water scarcity? To unravel this overarching question, I will delve into several sub-research questions, including: Which kind of water shortage has the biggest effect on crop growth?What is the role of the introduction of water scarcity response technologies in the crop yield and quality?Based on a thorough analysis of the literature, empirical research and real-life case studies, my research will help to develop a more detailed understanding of water scarcity and the ways to achieve sustainable solutions. Through the identification of practical strategies, the promotion of interdisciplinary dialogue and the promotion of meaningful action, I will be able to lay the foundation for a more resilient and water secure agriculture and our planet in general. As I begin this voyage of discovery and exploration, let us follow the footsteps of the past, get inspired by the dreams of the future, and set a direction towards a world where water shortage is not a coming disaster but a problem solved by collective wisdom and innovation.

## 2. [Literature Review]

The research on Key Performance Indicators (KPIs) in Vertical Farming Production Systems (VFPS) has been the key to the understanding of various aspects of the system performance. Nevertheless, these studies have mostly been divided into separate parts, concentrating on certain areas such as productivity, sustainability, or quality, instead of giving a complete picture of VFPS performance. The studies on productivity have mainly been focused on the aspects like the crop yield, growth rates and the resource utilization efficiency. These studies are designed to improve production output and at the same time reduce the resources used. Although productivity is important, the emphasis on it alone disregards other significant factors of VFPS performance like environmental impact and product quality (Fasciolo et al., 2024). Likewise, the studies that were made for sustainability have also looked at environmental impact, resource efficiency, and economic viability within VFPS. These actions are geared towards the creation of more ecological agricultural practices by cutting down on the use of resources and the environmental impact. Nevertheless, the sustainability research usually ignores the factors of productivity and product quality, which in turn, narrows the view on the VFPS performance. The quality-related studies, on the other hand, have been conducted to investigate the nutritional value, safety, and consumer preferences of VFPS products (Rome, 2022). These studies are the main reason for the safety and nutritional adequacy of vertical farming produce. Although, they may not be able to take into account the wider consequences of productivity and sustainability on the whole system performance. In summary, the disjointed nature of the previous research prevents the understanding of VFPS performance and thus the development of the standard evaluation methodologies is hindered. The absence of a well-structured system for the classification and assessment of KPIs across all aspects of life may lead to the inconsistencies in the measurement and interpretation. Thus, there is a great demand for research that combines productivity, sustainability and quality dimensions to provide a more complete evaluation of VFPS performance. These combined approaches can be used in the decision-making process, lead to the improvements, and thus, contribute to the development of vertical farming as a sustainable agricultural practice.

  The relevant literature for this study is the research articles, reviews, and theoretical frameworks dealing with the KPIs in the VFPS. Other researches have been conducted on different aspects of performance evaluation, such as productivity, sustainability, and quality. Productivity-related literature usually deals with crop yield, growth rates and the efficiency of resource utilization. Sustainability literature is a literature that studies the environmental impact, resource efficiency, and economic viability. Quality-related studies are the studies that deal with nutritional value, safety, and consumer preferences. Although each dimension has been studied separately, there is a need for research that combines all dimensions to give a full evaluation of the performance of VFPS.

  The synthesis of the previous literature shows the usual themes and trends in VFPS performance evaluation. The studies of productivity show that the maximum of crop yield and resource efficiency should be used to increase the production. These researches usually comprise of the examination of the new cultivation methods or the environmental conditions that can be used to improve the crop productivity. One example is, the attempts are made to enhance the hydroponic systems to the production of more crops with less resources. The other hand, the sustainability literature stresses the necessity of reducing the environmental impact and the resource consumption while keeping the economic situation stable. The studies usually focus on technological endeavors to cut down the environmental footprint, for example, the energy efficiency, the use of renewable energy sources, and the wastewater treatment and recycling.

Furthermore, the research on whether VFPS systems are socially acceptable and operated fairly is also very important.

Quality-related research in the field of Vertical Farming Production Systems (VFPS) is focused on the investigation of the nutritional value, safety and consumer satisfaction of the crops produced in these systems. To begin with, these studies are about the adherence to the food safety and quality standards which are the rules and regulations of the regulatory authorities. This means that the crops grown in vertical farming systems should be safe from contaminants such as pesticides, heavy metals, and pathogens and meet the rigorous safety criteria. Besides, the experts survey the cultivation environment to make sure that the soil and moisture conditions are suitable for the best plant growth and the nutrient uptake. In VFPS, where soil-based farming is not possible, the research is being done on the alternative ways of growing the plants like hydroponics, aeroponics, and aquaponics, which ensures the plants get the nutrients and hydration they need. Besides, the technical and production facets are thoroughly examined to satisfy the consumer's demands and preferences. The factors that contribute to the output quality are the variety of the crops that are selected, the cultivation methods that are used, the harvesting techniques that are employed, and the post-harvest handling procedures. For example, scientists may carry out a study on the use of certain plant types that are loved by consumers for their taste, look, or nutritional content. Besides, the research on the quality of VFPS also includes the activities that are aimed at the improvement of the sensory features of VFPS products so that the consumers will be more satisfied. This means, manipulating the factors like texture, flavor, aroma, and appearance of the crops through controlled environmental conditions and cultivation practices to get the best possible results. In general, the research in VFPS related to quality is designed to make sure that the crops produced are both safe and nutritious and at the same time they meet the consumer's preferences, thus, the marketability and the acceptance of the vertical farming produce are increased.

Nevertheless, the current research mostly concentrates on the individual performance aspects. This restricted method leads to a divided understanding of VFPS performance, since different studies focus on one particular aspect such as productivity, sustainability, or quality. Thus, there is not enough of a comprehensive analysis that includes all the performance dimensions, which in turn makes it difficult to understand the VFPS performance. Considering these limitations, future research should focus on the creation of the methods and the frameworks that can combine all the performance evaluation factors in a comprehensive way. Such techniques and frameworks would enable the comprehensive analysis of VFPS performance by including the various performance evaluation criteria. The adoption of such a complete method is the key to the accurate evaluation and the understanding of the performance of VFPS. Besides, the adoption of all-inclusive research methods would make it possible for the future studies to provide more detailed and meaningful results.

The gap in the literature is the absence of the research that covers all aspects of the performance evaluation in VFPS. Even though the studies on productivity, sustainability, or quality have been done before, there is still a need for research that will give a comprehensive evaluation of VFPS performance. The research aims to fill this gap by creating a thorough framework for the classification and assessment of KPIs in VFPS. Through the combination of the existing literature and the formulation of a systematic approach to the evaluation of VFPS performance, this study is going to give information about the many aspects of VFPS performance.

The theoretical foundation of this study is the multidimensional evaluation of VFPS performance. It encompasses three main dimensions: productivity, sustainability, and quality are the three main benefits of 3D printing. In each dimension, several concepts and variables are taken into account to give a complete evaluation of VFPS performance.

In the productivity aspect, the main idea is to evaluate the efficiency and effectiveness of crop production within VFPS. The main factors that are considered are the crop yield, the growth rates, the resource utilization efficiency, and the overall plant production. The connection that is supposed to be is that the higher productivity, shown by the increase of crop yield and the efficient use of resources, is the one that is related to the better performance of VFPS.

In the sustainability aspect, the attention is on the reduction of the environmental damage, the maximum use of the resources, and the economic feasibility (Benke & Tomkins, 2017). The crucial variables are the environmental input and output KPIs like energy consumption, greenhouse gas emissions, water use efficiency, and the economic indicators that are related to the cost-effectiveness and profitability. The anticipated connection is that the sustainable practices, which are the ones that cut down the resource consumption and the environmental impact, result in the more efficient and resilient VFPS operations.

In the quality aspect, the focus is on providing safe, nutritious, and high-quality products to the consumers. The notions like food safety standards, nutritional value, and consumer satisfaction are the main points. Variables are the ones that are related to food safety (e. g. the level of food contamination, the number of foodborne illnesses, etc. ). g. , nitrate content, natural toxic compounds), nutritional value (e. g. protein content, essential fatty acids, vitamins, minerals). g. , total soluble solids, fatty acid content), and consumer preferences (e. g. , taste, health benefits, and price). g. , taste, texture). The supposed connection is that the quality standards are high and that is the reason why the consumers are satisfied and trust the company, therefore, the market demand and the success of VFPS will be increased.

Through the combination of these dimensions and the study of their concepts and variables, the theoretical framework gives a systematic way to evaluate VFPS performance in a comprehensive manner. This model allows the researchers to evaluate the effectiveness and sustainability of VFPS operations, to find the areas for improvement and to advise the decision-making processes to optimize the system performance.

## 3. [Methodology & Data]

In this section, I will discuss the methodology and data collection process for my research, focusing on how I address my research questions: "What is the kind of water scarcity that has the most effect on the crop production?" and "How does the introduction of water scarcity response technologies affect the crop yield and quality?"

For the sake of my research questions, I use a mixture of both qualitative and quantitative methods to get a comprehensive answer. This method enables us to take into account the various aspects of the water scarcity's effect on crop production and to evaluate the water scarcity response technologies from a variety of viewpoints.

For the qualitative part of my study, I carry out the extensive reviews of the documents to study the qualitative effects of water scarcity on crop production. The review of documents includes the systematic examination of the literature, reports, and policy documents that are related to water scarcity, agricultural production, and technological interventions. By means of this in-depth qualitative analysis, I intend to acquire deep insights into the hidden forces which are the cause of the impact of water scarcity on crop yield and quality. At the start, the document review process is the one of the evaluation of the previous research and the theoretical frameworks in order to get the general idea of the topics that are related to the research. Through the careful examination of the collected literature, reports, and policy documents, I detect the key themes, patterns, and trends, which helps me to get a better understanding of the correlation between crop production and water scarcity. Besides, the review of the document is a way of getting different viewpoints and finding new ideas. In the process of integrating and comparing the theoretical models and the related studies that are present in the existing literature, I examine the various factors and the mechanisms through which water scarcity affects crop production. This all-round strategy makes the research more in-depth and wide-ranging, thus, it renders a complete analysis of the topic. Besides, the document review process also helps in giving the direction to the research and setting the basis for the further investigation. The theoretical frameworks and relevant topics found in the literature review act as the basis for the future steps and the focus of the research. To sum up, the qualitative methodology applied in this research is a good way of comprehending and evaluating the impact of water scarcity on crop production. The various materials collected from the documents are the reason why the research is wider and deeper at the same time, and thus, the researcher is able to understand the complex relationship between the crops production and water scarcity more clearly.

  Through the quantitative methodology, the researcher will be able to systematically and accurately collect and analyze the empirical data on water scarcity and its effect on crop production. Statistical methods are applied to study datasets with various indicators of water scarcity like Green Water Scarcity (GWS), Blue Water Scarcity (BWS), and Agricultural Economic Water Scarcity (AEWS) in order to find out patterns, correlations, and trends. The data collection process consists of the sourcing of data from the credible repositories, the governmental agencies and the academic institutions to make sure that the datasets used in the analysis are reliable and valid. In particular, statistical techniques like regression analysis are utilized to find the significant relations and trends in the dataset. Through the use of quantitative methods, the study aims to give concrete proof of the hypotheses regarding the unequal effects of water scarcity types and the efficiency of the response technologies on crop production. The main objective of the quantitative methodology is to assess the effect of water scarcity types and response technologies on crop production and to come up with statistically significant conclusions. These statistical methods help to check the hypotheses or claims, which in turn, gives the useful information for the formulation of the policies or strategies related to water scarcity or crop production.

  I apply statistical techniques to the data that is empirical and I quantify the relationships between the water scarcity types, the response technologies, and the crop production results. In particular, I use statistical methods such as regression analysis to find the significant links and the patterns in my data set. Through the use of quantitative methods, I want to give factual evidence that will support my hypotheses on the different effects of water scarcity types and the efficiency of response technologies on crop production.

  According to my literature review, I pinpoint the main concepts that are directly connected to my research queries, that is, water scarcity types, crop production outcomes, and the application of response technologies. For each construct, I set the particular ways to measure them in my study. For instance, water scarcity types are defined based on the duration of the scarcity periods (less than 1 month, more than 1 month), and the crop production outcomes include the yield quantity and the quality parameters (e. g. , the nutritional content, the market value).

The process of data collection for my research has various stages. Initially, I gathered a complete dataset which includes data on water scarcity events, agricultural practices, and crop production indicators. This dataset could be composed of data taken from the worldwide databases, remote sensing imagery, agricultural surveys, and experimental trials.

This study follows the ethical principles strictly from the data collection to the analysis and reporting. All information that is used for literature review and data analysis is from credible sources, thus, guaranteeing the transparency and the accuracy of the whole process . Besides, any possible biases or limitations that may have been encountered during the research process are reported honestly and openly (Singh & Chauhan, 2012). The study mainly uses data that are in accordance with the copyright regulations, and the source of the data is clearly attributed to protect the intellectual property rights of others. To keep the transparency and academic integrity, the study avoids distorting or manipulating the results and gives honest and objective analysis. This ethical practice improves the credibility and validity of the research and hence, it is the way to fulfill the ethical responsibilities with other researchers and the academic community.

My methodology is a mixture of qualitative document reviews and quantitative statistical analysis which is used to tackle my research questions in a comprehensive way (Fleetwood, 2024). Through the combination of different methods and data sources, I intend to offer a solid and detailed picture of the effects of water scarcity on the crop production and the effectiveness of the response technologies in the reduction of these effects. By conducting thorough data collection, analysis, and ethical studies, I aim to offer the field of agricultural water management and sustainability with valuable insights.

## 4. [Results of Analysis]

This study, which was about the effects of water shortage on aquaculture production and the possible responses, used a strict statistical analysis technique to analyze different types of water shortage. By numerical analysis, I sought to get the full picture of the impact of these factors on crop production. Through the analysis of the water scarcity types distribution, I was able to identify the complicated patterns and trends of the way each type affects the crop productivity. The statistical analysis gave me the chance to go into the data and to discover the hidden connections and the relations between the variables which are connected to water scarcity and crop production. I used the latest statistical methods to find the important trends and patterns which, in turn, gave me the knowledge about the complicated dynamics that are there in the activity. I mainly carried out a thorough analysis of the percentage distribution of water scarcity types in order to assess their effect on crop production during long periods of water scarcity (>1 month). This meant that they used statistical methods to calculate the number of cases of each water scarcity type and to evaluate the impact of it on the crop productivity for a long period of time. Through the application of statistical analysis techniques, I could get the important information from the data, thus, the data on the relationship between water scarcity and aquaculture production was revealed. Through this analytical method, I was able to reach solid conclusions and to come up with evidence-based recommendations for facing the water scarcity problems in aquaculture. In brief, the use of statistical analysis techniques in our study was a great step towards the improvement of the scientific validity of our work, thus, I was able to present comprehensive insights on the various aspects of water scarcity and its effects on crop production in aquaculture.

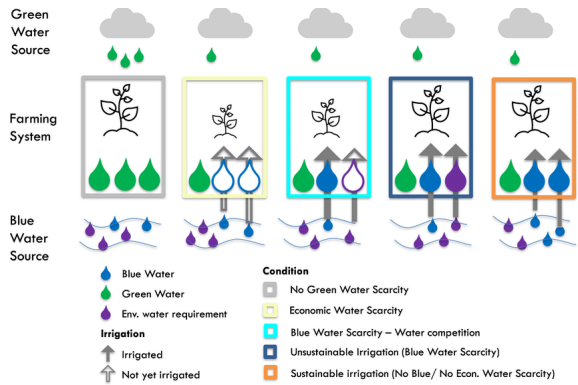

| Water scarcity type in global croplands | Time period in one year | |
|---|---|---|
| | <1 month | >1 month |
| | Percentage (%) | |
| GWS | 76 | 42 |
| BWS | 68 | 37 |
| AEWS | 43 | 86 |

Figure 1: Nature Communication          Figure 2: Water Scarcity Distribution In Global Croplands

    The table that shows the water scarcity in agricultural production that is analyzed in the text, makes it possible to understand the different effects of the water scarcity on the crop production. In this context, it's crucial to understand the definitions of the water scarcity types presented: GWS, BWS, and AEWS are the three types of intelligent systems (Ingrao et al., 2023). To start with, the GWS stands for the shortage of water in the soil that hinders the crop growth (Rosa et al., 2020). It mainly takes place because of the lack of precipitation which causes the soil to be dry and thus, the crops do not grow well especially during the critical growth stages. After that, Blue Water Scarcity (BWS) scarcity is the condition of a deficit of surface and groundwater resources, which are the basic needs for irrigation and other agricultural purposes (Liu et al., 2017). It usually comes from the overuse of water resources, pollution, and climate change that reduces the amount of water for farming. Furthermore, the Agricultural Economic Water Scarcity(AEWS) is a term that depicts the economic consequences of water scarcity on the agricultural areas (Chiarelli et al., 2022). It covers the elements that are related to the price of water, the possibility of water-saving technologies, and the economic feasibility of the agricultural methods in the water-scarce conditions. Every kind of water scarcity has its own specific effects on agricultural productivity, which are based on the combination of environmental and economic factors. AEWS is mainly the result of economic water scarcity, but BWS and GWS may have environmental factors as their cause such as climate variability and land management practices. The knowledge of these differences is necessary for the planning of the unique strategies to solve the water scarcity problems and to increase the agricultural resilience.

    The data clearly shows that AEWS is the most influencing factor on the croplands during the periods of prolonged scarcity (>1 month), affecting 86% of regions. This means that economic issues connected to water shortage are the main reason for the drop in crop production. AEWS is the representation of the economic limitations and problems of the agricultural regions which are the difficulties they face in getting and using the water resources properly. Thus, the information suggests that economic water scarcity, which is denoted by AEWS, is the most significant factor in crop production compared to other types of water scarcity, such as GWS and BWS.

The water scarcity response technologies are the ones that will be the main focus of the study on the influence of water scarcity on crop yield and quality. The analysis of the data can give the information on the effectiveness of these technologies in reducing the bad effects of water scarcity on crop production. For instance, if areas that are facing economic water scarcity have high levels of AEWS, and after they have adopted the water scarcity response technologies like vertical farming, they show the improvements in crop yield and quality metrics, it can be concluded that these technologies are useful to enhance the agricultural resilience to economic water scarcity (Vallino et al., 2020).

The researchers can find out the degree to which the water scarcity response technologies are able to improve the crop yield and quality in regions that are experiencing water scarcity by correlating the different types of water scarcity with the agricultural productivity and quality indicators in the regions that are using the water scarcity response technologies. The statistical analysis of such data gives researchers the chance to find the patterns, trends, and correlations, and thus, they can prove the results by the facts about the water scarcity types and the response technologies which are the factors of the crop production outcomes.

In conclusion, the statistical analysis of water scarcity data helps the researchers to answer the research questions about the impact of water scarcity on crop production and the influence of water scarcity response technologies (Dinar et al., 2019). Through the investigation of the distribution of the water scarcity types and their relationship with the agricultural outcomes, researchers can get the important information about the water scarcity and its effects on the crop production, thus, they can use the information for the decision-making and the policy formulation for the agricultural resilience to the water scarcity.

These findings show that the water scarcity issues are not just one-dimensional and the problem has both environmental and economic aspects, thus, the need to be aware of these aspects to increase agricultural resilience is stressed. Through the use of such insights, policymakers, researchers, and agricultural stakeholders can make evidence-based decisions, which, in turn, will help to manage water sustainably, to adopt the new technology, and to ensure food security in the areas of aquaculture production (Bunge et al., 2022).

The technological solutions to the water shortage in aquaculture production are very important, and the analytical findings prove that these solutions are the right ones. For instance, if water scarcity response technologies are introduced in the areas with high AEWS levels and as a result, there is an improvement in the crop production and quality indicators, it means that such technologies can effectively increase the agricultural resilience to the economic water scarcity (Dolan et al., 2021).

By looking at the different degrees of water scarcity types and their relations with the agricultural productivity and quality indicators, it is possible to evaluate how water scarcity response technologies are helping to improve crop production and quality in water-scarce conditions. The statistical analysis of the data enables the researchers to find the patterns, trends, and correlations, which in turn, provides the empirical evidence to the conclusions about the effect of water scarcity types and the effect of the response technologies.

Moreover, the effects of these analytical findings on agricultural policies, technological adoption, resource allocation, and research priorities can be different. Taking into account these findings, the legislators might have to include economic aspects in the water management plans to deal with the effect of AEWS. It might be suitable to adopt policies that will encourage the use of water-saving technologies and economic incentives to lessen the negative effects of economic water scarcity on aquaculture production. Thus, the primary focus should be on the use of water-saving technologies such as the efficient irrigation systems or hydroponic cultivation which will in turn, increase agricultural productivity and also, will help in the resilience to the economic water scarcity (Cairns, 2024). Finally, the resource allocation decisions based on the information are crucial in the areas that are suffering from severe water scarcity. The customized investments in infrastructure and technological interventions that are tailored to the water scarcity challenges can help to optimize resource utilization and to promote sustainable agricultural development.

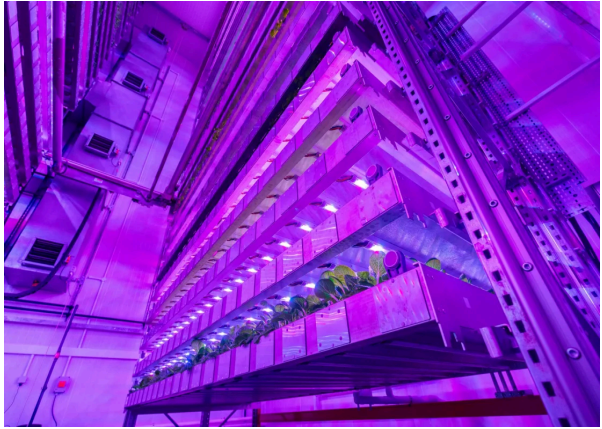 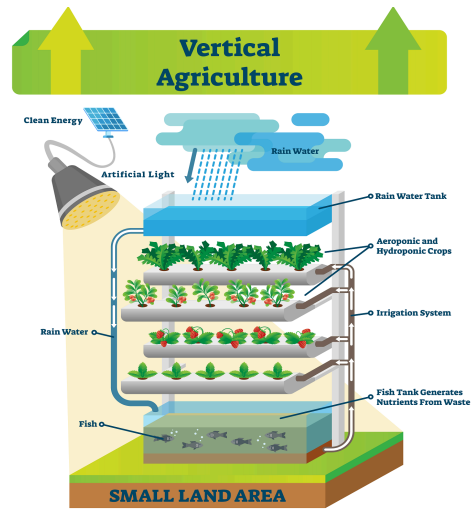

Image 1: Vertical Farming          Image 2: Process of Vertical Agriculture

Vertical farming is the technology that can solve the water scarcity problems that are mentioned in the analysis. It is an advanced agricultural technology that changes the old farming ways by growing the crops in the vertical layers (Oh & Lu, 2022). This novel technique enables the maximum utilization of land and the minimum of water usage, hence it is a very suitable solution to the water scarcity problems in agriculture. In contrast to traditional farming, which demands a lot of horizontal space, vertical farming uses agricultural technology to create the best conditions for the crops, which is controlled by the environment. Another important principle of vertical farming is its capability to produce more crops in a smaller area than conventional agriculture. This is the result of the crops being grown in the tower-like structures, which can be located in different places like shipping containers, warehouses, skyscrapers, and underground bomb shelters (Linnea, 2023). Vertical farming is a method of growing crops by stacking them vertically, thus, it makes it possible to grow a lot of crops in a small space and hence, it can be used in both urban and rural settings (Vejgaard, 2022). In the case of vertical farming, artificial lighting is of great importance in the process of giving the light energy that is needed for the growth of plants. LED lights, fluorescent grow lights, and high-pressure sodium lights (HPS) are the most popular methods of creating sunlight-like light and enabling photosynthesis (IMI, 2022). This means that the vertical farms can work independently of the seasonal variations in the sunlight, thus having a year-round production capability. Furthermore, vertical farming is based on other types of growing methods that do not use soil, such as aeroponics, hydroponics, and aquaponics. The plant roots are suspended in the air and are sprayed with a nutrient-rich solution, which is the way of getting the plants with the essential nutrients while at the same time, the water efficiency is maximized. Hydroponics is a technique of growing plants directly in a water-based solution that is full of nutrients, therefore, the soil is not needed and the water consumption is reduced (Rajendran et al., 2024). Different inert materials like coconut fiber, vermiculite, peat moss, perlite, and rockwool serve as a support for plant roots in the hydroponic systems. To sum up, aquaponics is a combination of aquaculture and hydroponics, with the plants and fish helping each other in a closed-loop system. To sum it up, vertical farming is a feasible and effective way to solve the water scarcity issues in agriculture. Thanks to the application of the latest technologies and the introduction of new farming techniques, vertical farming is able to enhance water efficiency, save resources and promote environmental sustainability. Vertical farming is a good method to increase food security, to manage the water resources sustainably, and to make agriculture more resilient in water-scarce areas (Wang et al., 2023). The research and innovation in vertical farming technology are the keys to the improvement of its potential and the widespread adoption of it as a cornerstone of the future agricultural systems.

Vertical farming becomes a versatile solution that can successfully solve the various types of water scarcity problems. The tool in question is especially effective in the fight against GWS due to its ability to control soil moisture levels precisely (AgTech, 2024). Through the provision of precise irrigation and nutrient delivery to plants, vertical farming reduces the chances of the soil moisture being too low that it will negatively affect the crop growth, hence it is a useful asset in the GWS regions where there is not enough rainfall. Besides, hydroponic systems of vertical farming greatly cut the water usage in contrast to the traditional soil-based agriculture methods, thus, it is the best option to fight Blue Water Scarcity (BWS). These systems are the most effective in transporting water and nutrients directly to the roots of the plants, thus making sure that the plants are well hydrated and at the same time, the water wastage is reduced. Therefore, vertical farming is a good way to solve the problems of BWS that are caused by the overexploitation and the contamination of the groundwater resources. Although AEWS is the cause of the economic problems, vertical farming's resource-efficient practices can be the solution to these problems. Through the enhancement of production per unit of water input, vertical farming makes the economic viability better, thus, it helps to solve the problems of AEWS such as the fluctuating water prices and the supply instability. To sum up, vertical farming is a versatile tool that can be used to deal with all the aspects of water scarcity, thus, it is a holistic solution for sustainable agricultural development. Its capacity to deal with GWS, BWS, and AEWS challenges shows its potential to increase the agricultural resilience and food security in water-stressed regions.

## 5. [Conclusion]

To sum up, this in-depth investigation of the connection between water scarcity and agricultural production reveals the importance of the issue of this critical global problem. A multi-dimensional analysis has been carried out which has helped to unravel the complex dynamics of water scarcity and its various types and their impacts on crop production. The results prove the deep impact of water scarcity on agricultural resilience, which, in turn, shows the necessity of innovative solutions to the problems of its negative effects.

Main points derived from the analysis show that the economic water scarcity (AEWS) is the most important factor that influences crop production. This economic constraint proves the significance of the use of technologies and practices that make water efficiency and economic viability in agriculture possible. Vertical farming is a viable way of overcoming the three types of water scarcity, which are Green Water Scarcity (GWS), Blue Water Scarcity (BWS) and AEWS. Its capacity to increase the efficiency of water use, to maximize the crop yield, and to reduce the environmental impact, makes vertical farming a fundamental element of sustainable agricultural development.

On the one hand, the understanding and solutions of water scarcity have been improved, but on the other hand, the limitations still remain. Our study is still inconclusive about the scalability and accessibility of the vertical farming technologies, and their socio-economic implications are still unclear. Subsequent research projects will have to investigate these aspects more thoroughly to enable the formulation of more inclusive and equitable water management strategies.
In the future, we will expand our approach to include more stakeholders and viewpoints, and this will lead to the interdisciplinary collaboration to deal with the complex situation of water scarcity and agriculture. Besides, the continuing actions will be directed to improving the techniques for the assessment of the effectiveness of the water scarcity response technologies, thus, the decision-making will be based on the solid and the scientific data.
Basically, this study is the trigger for the continued discussion and action on the building of resilience and sustainability in the agricultural systems against the background of water scarcity. Through the use of technological innovation, the process of interdisciplinary collaboration and the focus on the inclusive approach, we can set a course to the future of a water-secure agriculture and society.